\documentclass{pasj00}

\begin{document}
\SetRunningHead{Author(s) in page-head}{Running Head}
\Received{2002/10/31}
\Accepted{2002/10/31}

\title{The radiant of the Leonids meteor storm in 2001}

\author{Ken'ichi \textsc{Torii}  %
and Mitsumiro \textsc{Kohama}
}
\affil{Cosmic Radiation Laboratory, RIKEN,
2-1, Hirosawa, Wako 351-0198}
\email{torii@crab.riken.go.jp}

\author{Toshifumi \textsc{Yanagisawa}}
\affil{National Aerospace Laboratory of Japan, 7-44-1 Jindaiji Higashi-machi, Chofu 182-8522}\email{tyanagi@nal.go.jp}
\and
\author{Kouji {\sc Ohnishi}}
\affil{Nagano National College of Technology, 716 Tokuma, Nagano 381-8550}\email{ohnishi@ge.nagano-nct.ac.jp}
\affil{National Astronomical Observatory of Japan, 2-21-1 Osawa, Mitaka, Tokyo 181-8588}

\KeyWords{methods: data analysis, meteors, Solar system: general, comets: individual (55P/Tempel-Tuttle)} 

\maketitle

\begin{abstract}

 We have measured the radiant of the Leonids meteor storm in November
2001 by using new observational and analysis techniques. The radiant
was measured as the intersections of lines which were detected and
extrapolated from images obtained at a single observing site (Akeno
Observatory, Japan). The images were obtained by two sets of telephoto
lenses equipped with cooled CCD cameras. The measured radiant, (R.A.,
Dec.)=(154$^\circ$.35, 21$^\circ$.55) (J2000), is found to be in
reasonable agreement with the theoretical prediction by McNaught and
Asher (2001), which verifies their dust trail theory.

\end{abstract}

\section{Introduction}

 Recent progress of numerical celestial mechanics has made it possible
to accurately predict the occurrence time, position, and rate of
meteoric activities (e.g., \cite{mcnaught1999}; \cite{lyytinen};
\cite{mcnaught}).  These theories predict that meteor storms occur
when the Earth passes through the dense regions of meteoroids, or dust
trails. The dust tubes are produced along the trails of the comet near
perihelion passages and are kept in narrow tubes by combined effects
of gravitational perturbations from planets and solar radiation
pressure.  The development of dust trail theories is important not
only from celestial mechanics or interplanetary physics points of view
but also from planning the strategies for protecting artificial
satellites or manned missions against collisions of meteoroids (e.g.,
\cite{pawlowski}; \cite{brown}).

 Although these theories have successfully predicted the peak time of
meteoric activities, no extensive verifications have been made from
other aspects. Since the radiant of meteors is directly related to the
dust trails, the observational measurements of the radiant can be
critical test for the dust trail theories. For 2001 November activity
of Leonids meteor streams, McNaught and Asher (2001) predicted the presence of
two strong outbursts (ZHR$\sim$ 2000 and 8000) over East Asian longitudes each
of which is due to meteoroids released from the comet
55P/Tempel-Tuttle around its 1699 (9 revolutions ago) and 1866 
(4 revolutions ago) return, respectively.

We have thus made observations for verifying the dust trail theories
(\cite{ohnishi}; \cite{yanagisawa}). Our method differs from
conventional ones in several ways. We use cooled CCD cameras with
relatively narrow field of view optics. These instruments enable
accurate ($\leq$10 arcseconds) measurement of each meteor with
reference to background fixed stars. We have also employed a new image
processing technique (\cite{yanagisawa}\footnote{The technique is
patent pending.}) which effectively picks up faint line shapes of
meteors superposed on the background fixed stars. Our aim here is not
to measure the orbit of each meteor as conventionally studied but to
measure the apparent radiant point projected on the celestial
sphere. We therefore did not make the observations from multiple
observing sites. Two sky positions which subtend nearly right angle to
the predicted radiant were observed. Radiant is obtained as the
intersections of orthogonal lines detected from the two cameras.

\section{Observations}

 Observations were made on 2001 November 18 UT on the premises of
Akeno Observatory (35$^\circ$47$'$N, 138$^\circ$30$'$E, 900m altitude)
 of Institute for Cosmic Ray Research, University of Tokyo. We used unfiltered telephoto
lenses with the focal lengths of 180mm (Nikkor) and cooled CCD cameras
as summarized in table 1.  Camera 1 is Apogee's AP7p with the backside
illuminated CCD SITe SI-032AB (512$\times$512 pixels of
$24\mu$m$\times24\mu$m).  Camera 2 is Apogee's AP6E with the Kodak's
CCD KAF-1001E (1024$\times$1024 pixels of $24\mu$m$\times24\mu$m). The
exposures were continuously made with integration times of
20-s. Readout times are 10-s and 3-s for the cameras 1 and 2,
respectively, which result in $\sim 33$\% and $\sim 13$\% of dead time
in the observation.  The pixel scales were $28''$/pixel and
$27''$/pixel for the cameras 1 and 2, respectively. Limiting
magnitudes for background fixed stars were $\sim 12-13$ mag for a
single frame.  These cameras were placed on an equatorial mount and
tracked at the sidereal rate. The pointing positions (centers of the
field of view), as summarized in table 1, subtended $\sim90^\circ$ to
the expected radiant.  Camera 1 pointed at a position about
$-20^\circ$ (minus sign means westward) away from the radiant along
right ascension while the camera 2 pointed at a position about
$+40^\circ$ (plus sign means northward) along the declination. The position angles of the cameras were slightly rotated from the
north so that the line shapes of the meteors do not become parallel to
the column or row of the CCD pixels. This setting reduces the false
events due to instrumental effects (e.g., column defects of the chip). Table 2
shows the start times of exposures for the first and last frames for
the two cameras as well as the total number of frames.  The weather
condition was very good and no significant clouds bothered our
abbreviations.

\begin{figure}
  \begin{center}
    \FigureFile(80mm,80mm){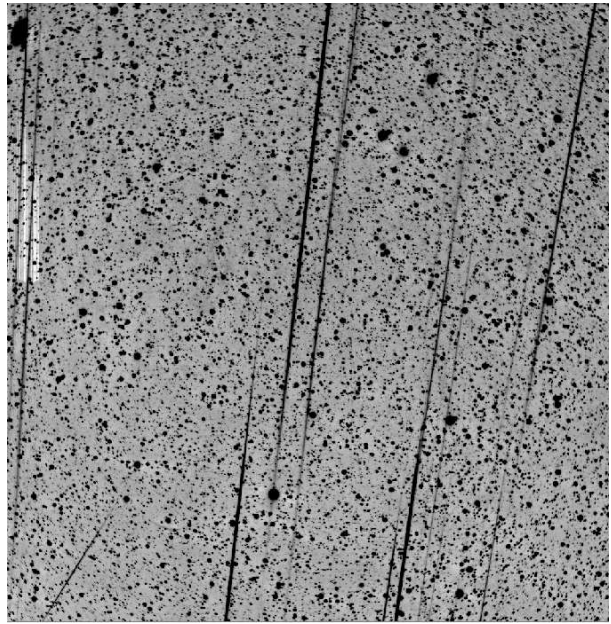}
  \end{center}
  \caption{Sample bright meteors as observed by camera 2. 
}\label{fig1}
\end{figure}

 We detect linear shapes of meteors from the obtained images and
extrapolate the lines toward the radiant. The radiant is thus
determined as intersections of many lines detected by the two cameras.
Systematic errors for the determination of lines come from several
factors. First, the fixed stars revolve $15\,T\, cos(\delta)$
arcseconds during the integration time of $T\, {\rm [s]}$, which leads
to uniform error within $\pm 2'.5$ for the current observation
($T=20\, {\rm [s]}$).
 In the current configuration, an error in the angle determination of a line
corresponding to 1 pixel is magnified to $1'.7$ and $3'.4$ at the
radiant. These two factors are combined to make total systematic error
of $\sim 4'$.  This value is marginally smaller than the expected
separation ($\sim 0^\circ.1$) of the two radiants corresponding to the
dust tubes of 4-revolutions and 9-revolutions ago (\cite{mcnaught}).
Figure \ref{fig1} shows some bright meteors as observed by the current
system. This image was created by stacking 65 frames from camera 2.

\begin{longtable}{cccc}
\caption{Observing Instruments.}
\label{tab:first}
\hline\hline
Camera & Optics & Field of view & Center of the Field (J2000)\\ \hline
\endhead
\endlastfoot
1 & 180mm f/2.8 & 3.9$^\circ$$\times$$3.9^\circ$ & (08 48 25, +19 21 38) \\ 
2 & 180mm f/2.8 & 7.8$^\circ$$\times$$7.8^\circ$ & (09 23 00, +65 06 02) \\  \hline 
\end{longtable}

\begin{longtable}{ccccc}
  \caption{Observation Log.}\label{tab:second}
\hline\hline
Camera & Start time [UT] & End time [UT] & Exposure [s] & Total number of frames\\ \hline
\endhead
\endlastfoot
1 & 2001 November 18, 15:11:43 & 2001 November 18, 20:52:31 & 20 & 671 \\ 
2 & 2001 November 18, 16:34:38 & 2001 November 18, 20:21:04 & 20 & 601 \\  \hline 
\end{longtable}

\section{Analyses}

The data were stored in the local hard disk and later processed
offline. After dark frame subtraction and flat fielding, astrometric
measurement was made for each frame with reference to USNO-A 2.0
catalog (\cite{monet}). The softwares
PIXY\footnote{http://www.aerith.net/misao/pixy1/} and imwcs
\footnote{http://tdc-www.harvard.edu/software/wcstools/imwcs/} were
used and the field center and the rotation angle were determined for
each frame. The accuracy of the astrometry (field center) is typically
better than the pixel scale 
and that
for the position angle is typically better than $\simeq 0.01^\circ$.

 We have used the new method (\cite{yanagisawa}) for detecting meteors
 from the observed frames. This technique was originally developed for
 detecting trails of space debris or artificial satellites from CCD
 images while it can be generally used for detecting line shapes on
 two dimensional images in the presence of point-like backgrounds.  The
 details of the algorithm is described in \cite{yanagisawa} and the
 method is briefly explained here.  Each image is rotated around its
 center by a trial angle $\theta_{rot}$. Then the central square
 region is extracted and the median values of each row is calculated
 and stored. In the absence of a line in the image, the median values
 are randomly distributed as background levels. If a line is present
 and the rotation correctly puts the line along the row, the median
 value becomes higher than those of adjacent rows, due to the
 systematic shift in the distribution of the pixel values toward
 higher side. Since the presence of point-like stellar images does not
 systematically shift the distribution, this technique effectively
 picks up line shapes from the background stars and noise
 fluctuations. For detecting a line of unknown angle, we make trials
 with different rotation angles $\theta_{rot}$ with small steps.  In
 the current analysis, we subtracted the $(i-1)$-th image (image of
 the previous exposure) from the $i$-th image so that the effects of
 fixed stars and small imperfection of flat fielding are
 further reduced.

\subsubsection{Results}

 We have examined the appropriate thresholds to discern real events
 (lines) from background fluctuations. To do this, we have created histograms
 of rotation angles at which the lines were detected. For bright real
 events, the angles are concentrated toward the radiant, while the
 background events are uniformly distributed. We set the thresholds so
 that the signal to noise ratios are more than 2 in the histogram.
Consequently, the
 limiting magnitude of  the current observation is estimated to $\sim
 7$ mag for meteors by the cross calibration with the wide-field TV
 observation at the same site. 

 As the results of the line detection analyses, we detected 9 and 80
lines from the camera 1 and 2. The small number of detected line for
the camera 1 is partly due to the presence of a stellar cluster (M~44
= NGC~2632) and a bright star ($\delta$ Cnc, $\sim$3.9 mag) within the
field of view which made background higher than that for the camera 2.

\begin{figure}[ph]
  \begin{center}
    \FigureFile(80mm,80mm){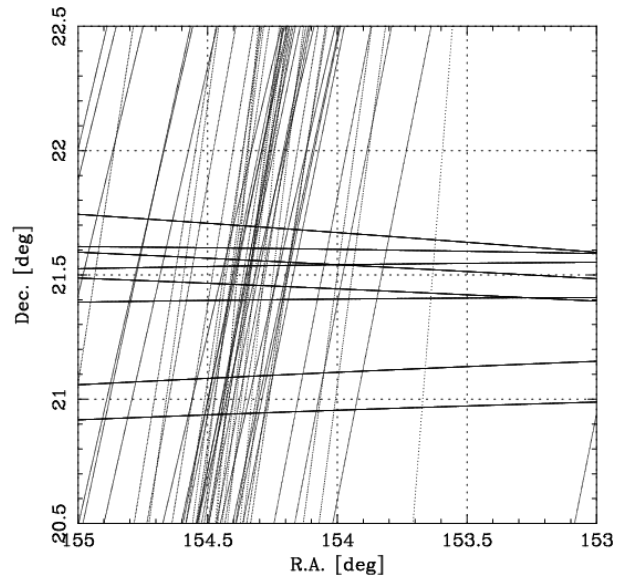}
  \end{center}
\caption{The extrapolated lines are shown at around the predicted
radiant. Horizontal (east-west) and vertical (north-south) lines show those from the camera 1 and 2,
respectively. These lines are obtained by extending the originally detected
lines back toward the radiant.}
\label{fig2}
\end{figure}

\begin{figure}[ph]
  \begin{center}
    \FigureFile(71mm,71mm){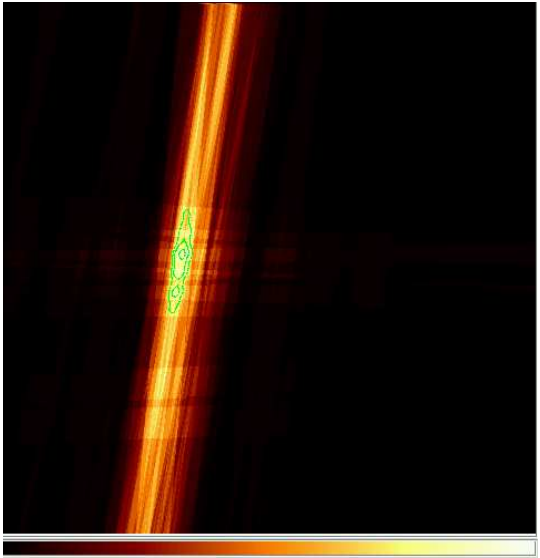}
  \end{center}
  \caption{The density of concentration of the extrapolated lines are
  shown at around the expected radiant. The field of view is the same as that for figure 2. The contour levels show the concentration of 19, 20, and 21 lines.}
\label{fig3}
\end{figure}

Based on the astrometry of fixed stars, we have converted the
positions of each line to the celestial coordinates (right ascension
and declination) and extrapolated (extended) back to their origin on the
spherical coordinates.
The apparent
position of radiant moves 
due to the combined effects of diurnal aberration
and zenithal attraction. These effects smear out the apparent
radiant in a short time of interval and makes it difficult to resolve
the radiant structures.  We therefore calculated the shift as a function of time and corrected the positions of each line to cancel the effect.
The reference time for this
correction was set to November 18 18:13 UT which was the predicted
peak time for the 4-rev trail encounter (\cite{mcnaught}).  This
procedure makes it possible to combine data of long duration to
improve the statistics.
The result is shown in figure \ref{fig2} and \ref{fig3}. 
The lines are distributed around the
expected position while we can clearly see the concentration at around
($\alpha$, $\delta$)=(154$^\circ$.35, 21$^\circ$.55).

\begin{figure}[ph]
  \begin{center}
    \FigureFile(80mm,80mm){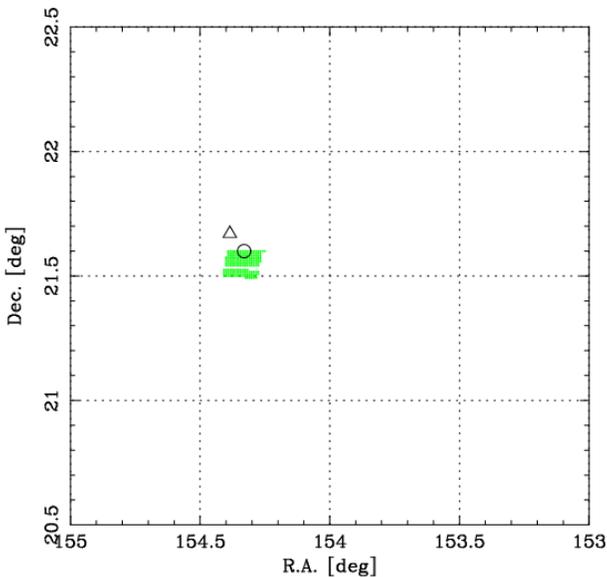}
  \end{center}
  \caption{The measured radiant structure is shown by filled green dots. The
  black  circle and triangle show the predicted radiant in Tokyo at 2001 November 18 18:13 UT, resulting from
  the dust ejection at perihelion minus 50 days for the 4-revolution
  and 9-revolution trails (Adopted from Figure 4 and Table
  2 of \cite{mcnaught} ). }
\label{fig4}
\end{figure}

To clearly see the radiant, we have examined the concentration of the
lines in the following way. For each grid point near the radiant, we
calculated the distance between the position and the lines. If the
number of lines within the threshold distance $r_{th}$ is 
more than the threshold numbers, that positions is considered as the
correct radiant. This procedure has thus three free parameters,
$r_{th}$, $n_1$, and $n_2$ which are not given apriori. We use
$r_{th}\leq 4'$, $n_1 \geq 3$, and $n_2\geq15$. This value of $r_{th}$ is chosen
so that the value is comparable to the systematic error as estimated
above. The values of $n_1$ and $n_2$ are determined by the number of
detected lines. Particularly, the value of $n_1$ had to be set to as
small as 3 due to the small number of detected lines from the camera
1.  The value of $n_2/n_1$ may be reasonable, taking into account the
effective area and exposures of the two cameras.
 Figure \ref{fig4} shows the radiant structure as finally
 determined. 

The position derived in the current work is in reasonable
agreement with the theoretical prediction of \cite{mcnaught}. The
measured position in declination looks displaced by $\sim -0^\circ.1$
from that predicted (Figure \ref{fig4}). However, we may not
conclude that they are inconsistent, taking into account the small
number of east-west lines used to constrain the declination. The
better constrained right ascention is consistent with the prediction
both in the central position and the extension. Although we expected to
resolve the two radiants from the two dust trails, they
could not be resolved partly due to the small number of the east-west
lines and partly due to the relatively large systematic error of the
current study. We find weak evidence of shift of right ascension from
RA$\sim 154^\circ.4$ to RA$\sim 154^\circ.3$ by the time resolved
analysis. Although it is as expected from the peak times of the two
dust trails,
limited statistics does not allow us to conclude that they
unambiguously come from the two distinct points or from relatively
extended ($\sim 0.^\circ1$) region.

The measured position is also found to be in reasonable
 agreement with the complementary result by using longer (530~mm) focal
 length optics at the same observing site (\cite{yanagisawa2}). Since
 all the hardwares and softwares are independent between the two
 works, this agreement ensures the adequacy of the whole procedures of the
two analyses.

\section{Conclusions}

 We have shown that the new method
presented here can be a powerful diagnostic tool for studying the
radiant structure of meteor storms. The measured position is found to
be in reasonable agreement with the prediction of \cite{mcnaught}
based on their dust trail theory.

In 2002, Leonids meteor storm is expected to be observed in North
America and in Europe. The observation and analysis method presented
herein will be useful to further resolve the profile structure of the
dust tubes. Cameras with large area, fast read-out CCDs such as ROTSE
(\cite{akerlof}), LOTIS (\cite{park}),and RAPTOR (\cite{borozdin}, \cite{vestrand})
may be most useful for detecting a large number of meteors in a short
time. If enough number of meteors could be detected in a short time,
the three dimensional (time resolved) structure of dust tubes may be
obtained with the current method. 


\section*{Acknowledgment}

The authors are grateful to Prof. Masahiro Teshima and all the staffs
of Akeno Observatory of University of Tokyo for their kind support
during our observations.



\end{document}